\documentclass[twocolumn,aps,floats,floatfix,nofootinbib,prl,superscriptaddress,tightenlines]{revtex4}
\usepackage{graphicx} \usepackage{bm} \usepackage{epsfig}
\usepackage{pstricks}

\newcommand{\nn}{\nonumber}

\newcommand{\beq}{\begin{equation}}
\newcommand{\eeq}{\end{equation}}
\newcommand{\bea}{\begin{eqnarray}}
\newcommand{\eea}{\end{eqnarray}}

\begin{document}
\title{The Hyperfine Einstein-Infeld-Hoffmann Potential}
\author{Rafael A. Porto}
\address{Carnegie Mellon University
Dept. of Physics, \\
Pittsburgh PA  15213, USA}
\author{Ira Z. Rothstein}
\address{Carnegie Mellon University
Dept. of Physics, \\
Pittsburgh PA  15213, USA}

\begin{abstract}
We use recently developed effective field theory techniques
to calculate the third order post-Newtonian correction to
the spin-spin potential between two spinning objects. This
correction represents the first contribution to the spin-spin
interaction due to the non-linear nature of general relativity and
will play an important role in forthcoming gravity wave
experiments.
\end{abstract}

\maketitle

The problem of motion in general relativity has shed its seeming
academic nature due to pending gravity wave experiments. Both LIGO
\cite{ligo} and LISA \cite{LISA} expect to detect radiation from
inspiralling binaries, and thus building templates for these
events has become increasingly important. While for late stages of
the inspirals numerical techniques are needed, for the early
stages one may calculate in the Post-Newtonian approximation (PN)
for small velocities. In relativistic cases one may also calculate analytically when deviations
from the Schwarzschild geometry are small. A complicating factor
in the building of these templates is the fact that there are
multiple scales involved. In particular, the finite size of the
object, leads to tidal deformations and dissipation which can then
in turn affect potentials as well as radiation. These effects make
an exact analytical solution intractable. One might hope that
systematically expanding around the point particle approximation
would lead to a controlled calculational scheme. This is in fact
what one can do by using techniques developed for effective field
theories \cite{nrgr1,nrgr2,TASI}. While these theories  are typically
applied to  quantum field theories involving multiple scales, we will
be applying them here to a purely classical problem. The classical
problem shares many of the same calculational hurdles as the
quantum problem. In particular, because we are expanding around
the point particle approximation, the calculation of potentials,
as well as the radiated power loss, entails regularizing
divergences. However, since in the EFT we work at the level of the
action, these divergences are simply renormalized\cite{nrgr1}, and for certain
higher dimensional terms in the action lead to a classical
renormalization group trajectory\cite{goldbergerwise}. The EFT also allows one to power
count in a very natural way. Each term in the action has a
definite scaling in the expansion parameter and thus we may
determine the size of a given effect simply by reading off the
scaling  of the relevant terms in the action.

In the last few years the 3PN ($(O(v^6)$) potentials for
non-spinning objects have been computed \cite{blanchet}, but the
case of spinning bodies has not reached that level of accuracy.
The 2.5PN spin-orbit potential was calculated in \cite{owen}
within a point particle approximation and Hadamard regularization.
The leading order spin potentials were obtained in \cite{BH}, but
the spin-spin piece has yet to be computed to 3PN. By using the
techniques developed in \cite{nrgr1} we may avoid any of the pitfalls of the point particle approximation and tame divergences as
well as finite size effects. Following the literature, in the
spinning case when we use the term 3PN, we mean  order $v^6
v_{rot}^2$, where $v_{rot}$ is the rotational velocity which is
taken to be order one, in the ``maximally rotating case" and
$\epsilon v$ in the ``co-rotating case'', where $\epsilon \equiv
R/r$ ($\sim v^2$ for compact objects), and $R,r$ are the radius of
the object and orbit respectively.

In this letter we will be applying these EFT ideas, extended to
spinning particles in \cite{nrgr3}, to calculate the
first-nonlinear correction to the spin-spin potential for compact
binaries which arises at 3PN \footnote{The results can be easily
extended to non-compact objects. In the latter the leading and
subleading contributions to the spin-spin potential scale as
$\epsilon^2v_{rot}^2$ and $\epsilon^2v_{rot}^2 v^2$
respectively.}. This correction is the ``hyper-fine'' analog of
the Einstein-Infeld-Hoffmann potential calculated more than
seventy years ago \cite{EIH}. While the high order of this contribution
makes it appear to be of only academic interest, such accuracy may indeed be necessary
for successful matching \cite{v6rational} due to the fact that
inspirals are tracked over long periods of time.

As explained in \cite{nrgr1}, the EFT approach proceeds in two
stages. First one matches the full theory of an extended object
interacting with gravity onto a world-line action treating each
object in isolation. This action consists of a tower of world-line
operators, with the coefficients of higher dimensional operators
encapsulating the true finite size nature of the underlying
particle. These coefficients can be fixed by a matching procedure
as discussed for the case of absorption in \cite{nrgr2}. At the
order we are working the world line action necessary for our
calculation is given by \bea \label{fullaction}
S=&-&\sum_a \int m_a ds_a + \int \frac{1}{2}S_a^{IJ}\Omega^a_{IJ}(x_a) d\lambda_a \nonumber \\
&-&M_{pl}^2 \int d^4x \left(2\sqrt{g} R+(\partial^\nu h_{\mu
\nu}-\frac{1}{2}\partial_\mu h^\alpha_\alpha)^2\right) \eea where
$s_a$ and $\lambda_a$  are the proper length  and  affine parameters for
the $a$'th worldline respectively. The last term is the gauge fixing term,
which corresponds to the harmonic gauge used in all our
calculations. In \cite{nrgr3}, following the classic work of Regge
and Hanson \cite{regge} for a relativistic top in flat space, the
spin effects are included in a generally covariant fashion by
introducing the vierbein degrees of freedom $e^\mu_I(\lambda_a)$
on the worldline, which relate the local co-rotating frame to the
global frame. The generalized angular velocity is then given by
(for each particle) \beq e^{J \nu}\frac{D}{d\lambda} e^I_\nu =
\Omega^{IJ} \eeq and the spin $S_{IJ}$ is the variable conjugate
to $\Omega_{IJ}$. Operators describing finite size effects have
been left out as they are subleading corrections to the spin-spin
effects we are interested here \cite{nrgr1,nrgr3}. We will come
back to this point later on. The form of the world-line action
($\ref{fullaction}$) is fixed by reparametrization invariance,
which allows us to freely choose the worldline parameter. A wise
choice is thus $\lambda=x^0$, which will directly lead us to the
effective action in the PN frame. Therefore, throughout this
letter $v^\mu \equiv (1,v^j \equiv \frac{dx^j}{dx^0})$.

We next match onto a theory of potentials. This is accomplished by
first expanding the action around the flat space limit, i.e \bea
\label{linearize}
g_{\mu\nu}&=&\eta_{\mu \nu}+h_{\mu \nu} \nonumber\\
e^J_\mu&=&\Lambda^J_\mu+ \frac{h_\mu^{\nu}\Lambda_{\nu
}^J}{2M_{pl}}- \frac{h_\mu^\rho h_{\rho \sigma} \Lambda^{\sigma
I}}{8M_{pl}^2}+...\eea where $\Lambda$ is the Lorentz
transformation that relates the local co-rotating and global
frames in the flat space limit.

Using the power counting rules developed in \cite{nrgr1,nrgr3} we
expand each term in powers of the relative velocity such that each
term in the action scales homogeneously in $v$. This allows us to
calculate potentials to arbitrary order in $v$ by drawing all
possible Feynman diagrams involving a fixed set of vertices.
Before specifying the action, we must choose a gauge for the spin
degrees of freedom as well as the tetrad. A common and convenient
choice are the Newton-Wigner coordinates \bea
e^\mu_0&=&\frac{p^\mu}{m} \nonumber \\
mS^{0\mu}&=& S^{\mu\nu}p_\nu \to S^{i0}=\frac{1}{2}v^j
S^{ij}+O(v^4),\eea in which the position and momenta are treated
as canonical variables and the spin obeys the traditional angular
momentum algebra \cite{regge}. These constraints, which are not
all linearly independent, eliminate the unphysical degrees of
freedom. Results in this gauge can be related to other choices via
the appropriate coordinate transformations \cite{will}.

Given this gauge choice, we expand out the action yielding to the
leading order spin graviton coupling \beq \label{lo}
L_2=\frac{1}{2M_{pl}}S^{ij} h^{0i,j}, \eeq which arises from
expanding $\Omega$ in terms of the tetrad and using
(\ref{linearize}). Given that the spin scale as \cite{nrgr3} \beq
S\propto \frac{m^2}{M_{pl^2}}v_{rot}, \eeq we find that this
leading order interaction goes like $v^2 v_{rot}$. Thus the
leading order spin-spin potential follows from the
Feynman diagram with two of these leading order interactions  and
thus will scale as $v^4 v_{rot}^2$. It is given by
a one graviton exchange diagram leading to
\beq
V^{ss}_{2PN}=-\frac{G_N}{r^3}(\vec S_1\cdot \vec S_2-3\frac{\vec r
\cdot \vec S_1 \vec r \cdot \vec S_2}{r^2}), \eeq which agrees
with the previously derived results \cite{BH}. We will use
throughout the paper $S^{ij} \equiv \epsilon^{ijk}S^{k}$ (for
different definitions see \cite{regge,owen}), and the standard
notation labelling the post-Newtonian order by the power of the
orbital velocity.

To work to next order we must consider vertices involving spin to
order $v^3 v_{rot}$. At this order we have \beq L_3
= \frac{1}{2M_{pl}}S^{j k}v^ih^{ik,j}+\frac{1}{4M_{pl}}v^lS^{j l}h^{00,j}\nn \\
\eeq while at order  $v^4 v_{rot}$  we must include \bea L_4 &=&
\frac{1}{4M_{pl}}v^lS^{lj} v^ih^{i0,j}+ \frac{1}{4M_{pl}}v^lS^{lj}
h^{0j,0}.\nn \eea These terms arise by keeping higher order pieces
in the spin matrix (i.e. $S_{0i}$) as well as time derivatives
which are down by $v$ relative to spatial derivatives. Since we
are calculating to order $v^6v_{rot}^2$ (3PN) we must include
diagrams with one insertion of $L_4$ and a leading order vertex
$L_2$, as well as those diagrams with two insertions of $L_3$.
These diagrams are depicted in Figures (1a) and (1b). There is a
further contribution from one graviton exchange which arises from
the first correction to the graviton Green's function \beq
G(p)\propto \frac{1}{p_0^2-\vec p^2}\approx -\frac{1}{\vec
p^2}-\frac{p_0^2}{\vec p^4}+.~.~. \eeq which reflects the first
deviation from instantaneity in the exchange. This contribution is
depicted in Figure (1c) and,  more formally, arise from new
vertices in the action \cite{nrgr1}.
\begin{figure}[t!]
\scalebox{0.54}{\includegraphics{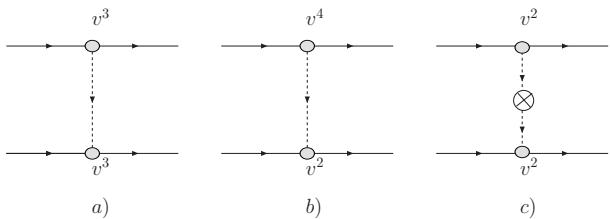}}\vskip-0.3cm
\caption[1]{ Diagrams contributing to 3PN order which do not
involve non-linear interactions. The blow represents a spin
insertion and the cross corresponds to a propagator
correction.}\label{ss3}
\end{figure}

At order 3PN we must also include diagrams with double graviton
exchange (note that we need not include diagrams which can be
disconnected by cutting the world-line, as these are iterations of
connected diagrams which are automatically resummed when solving
the equations of motion). Terms quadratic in the metric arise from
the higher order terms in the vierbein and the connection. After
expanding we find that the relevant pieces of the Lagrangian are
\beq L_{h^2}=  \frac{1}{4M_{pl}^2}S^{ij}( h^{i k} (h^{k 0,j}-
h^{j0,k})-
 h^{i 0} h^{0 0,j}).
\eeq These terms scale as $v^4v_{rot}$ thus, we must include all
diagrams for which a leading order spin vertex ($v^2v_{rot}$) and
a leading order mass insertion from the non-spinning part of the
action \beq L_0=-\frac{m}{2M_{pl}}h_{00},\eeq are contracted. The
resulting diagram is shown in Figure (2b). Finally we must include
the contribution from those diagrams including the three graviton
vertex which scales as $v^2$ \cite{nrgr1}. This vertex is too
voluminous to include here, and is best, and quite simply, handled
using a symbolic manipulation program
%\cite{avail}.
 All such
diagrams resulting from this vertex, two leading order spin
insertions and one leading order mass insertion must be included
(see Figure (2a)).

\begin{figure}[t!]
\scalebox{0.6}{\includegraphics{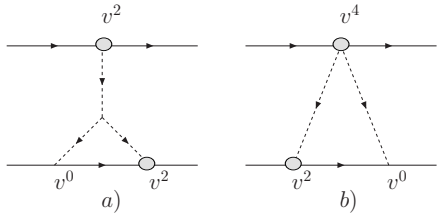}}\vskip-0.3cm
\caption[1]{Non-linear contributions to the 3PN spin-spin potential.}\label{ss2}
\end{figure}

Combining all of the contributions leads to the potential

\begin{widetext}
\begin{eqnarray}
V^{ss}_{3PN} &=& -\frac{G_N}{2r^3}\left[ \vec{S}_1\cdot
\vec{S}_2\left({3\over2}\vec{v}_1\cdot\vec{v}_2-3\vec{v}_1\cdot{\vec{n}}\vec{v}_2\cdot{\vec{n}}
-\left(\vec{v}_1^2+\vec{v}_2^2\right)\right)
-\vec{S}_1\cdot\vec{v}_1\vec{S}_2\cdot\vec{v}_2-\frac{3}{2}\vec{S}_1\cdot\vec{v}_2\vec{S}_2\cdot\vec{v}_1+
\vec{S}_1\cdot\vec{v}_2\vec{S}_2\cdot\vec{v}_2\right.\nonumber\\
&+& \vec{S}_2\cdot\vec{v}_1\vec{S}_1\cdot\vec{v}_1+
3\vec{S}_1\cdot\vec{n}\vec{S}_2\cdot\vec{n}
\left(\vec{v}_1\cdot\vec{v}_2+5\vec{v}_1\cdot\vec{n}\vec{v}_2\cdot\vec{n}\right)
-3\vec{S}_1\cdot\vec{v}_1\vec{S}_2\cdot\vec{n}\vec{v}_2\cdot\vec{n}-
3\vec{S}_2\cdot\vec{v}_2\vec{S}_1\cdot\vec{n}\vec{v}_1\cdot\vec{n}
\nonumber\\
&+& \left.
3(\vec{v}_2\times\vec{S}_1)\cdot\vec{n}(\vec{v}_2\times\vec{S}_2)\cdot\vec{n}
+3(\vec{v}_1\times\vec{S}_1)\cdot\vec{n}(\vec{v}_1\times\vec{S}_2)\cdot\vec{n}-
\frac{3}{2}(\vec{v}_1\times\vec{S}_1)\cdot\vec{n}(\vec{v}_2\times\vec{S}_2)\cdot\vec{n}\right.\nonumber\\
&-& \left.
6(\vec{v}_1\times\vec{S}_2)\cdot\vec{n}(\vec{v}_2\times\vec{S}_1)\cdot\vec{n}
\right]+\frac{3G^2_N(m_1+m_2)}{r^4}\left(\vec{S}_1\cdot\vec{S}_2-3\vec{S}_1\cdot\vec{n}\vec{S}_2\cdot\vec{n}\right),
\end{eqnarray}
\end{widetext}
with $\vec{n}\equiv \frac{\vec{r}}{r}$. The equations of motion
simply follow from the usual Hamiltonian procedure.

The leading order finite size contribution arise from so-called
self induced effects. That is, corrections to sphericity arising
from the non-vanishing quadrupole moments induced by rotation.
These effects are encapsulated by the world-line operator \beq
L_{ES^2} \equiv \frac{C}{2mM_{pl}}\frac{E_{\mu \nu}}{\sqrt{u^2}}
S^{\mu}_\rho S^{\rho \nu},\label{s2} \eeq where $C$ is a constant
which is determined by the nature of the object and $E_{\mu\nu}$
is the electric component of the Weyl tensor. In the case of a
rotating black hole $C=1$, and this term represents the
non-vanishing quadrupole moment of the Kerr solution
\cite{thorne}. For neutron stars $C$ ranges between $4$ and $8$
depending on the equation of state of the neutron star matter
\cite{poisson}. This operator scales as $v^4 v_{rot}^2$ at leading
order, and it is easy to show that it gives rise to a 2PN
(quadrupole-monopole) spin-orbit contribution
\cite{poisson2,nrgr3}, \beq
V^{so}_{2PN}=-\frac{C G_Nm_2}{2m_1r^3}\left(\vec
S_1\cdot\vec{S_1}-3(\vec n \cdot \vec S_1)^2\right)+ \; 1
\leftrightarrow 2, \eeq so one might then think that it will also
contribute at 3PN in the spin-spin sector, since naively one could
contract (\ref{s2}) with a leading order spin interaction
(\ref{lo}). However, this contribution vanishes, and the leading
order finite size effects in the spin-spin potential show up at
3.5PN. There is nevertheless a finite size correction in the
spin-orbit interaction at 3PN coming from diagrams where
(\ref{s2}) is contracted with subleading mass insertions as well
as corrections to instantaneity and non linear effects. We will
report on these results in a forthcoming paper.

As discussed in \cite{nrgr1} the inclusion of radiation into this
formalism is accomplished by working in a  background field. In
the  background field gauge, after multipole expanding the field
to keep manifest power counting \cite{grinroth}, and integrating
over (i.e. solving the equations of motion) the potential
gravitons, one generates an effective action for the sources and
the radiation graviton  $\Gamma[x,\dot x, h]$ which is invariant
under small diffeomorphisms. The power loss due to radiation then
follows by calculating  the imaginary parts of all vacuum diagrams
(i.e. no external gravitons), whereas the real part introduces
radiation reaction effects into the equations of motion for the
sources. Including spin effects in radiation follows in a
similarly straightforward manner by simply including vertices in
which the spin couples to the background field. We have calculated
the effects of spin on radiation to 2PN and find agreement with
those in the literature \cite{spinrad}. These results along with
new results for the 3PN contributions to radiation and
quasicircular orbits with the inclusion of spin, will follow in a
subsequent publication.

Finally, it is hopefully clear that there are no stumbling blocks
in going beyond 3PN in our formalism. It is simply a matter of
bookkeeping. Since every term in the action scales homogeneously
in the velocity this is not a difficult task. One simply draws all
possible diagrams such that the net scaling of all the vertices is
of the desired accuracy. For instance, at next order one would
need the four-graviton vertex as well as the higher order vertices
with and without spin and a contact term involving one source and
three graviton lines. Divergences are not an issue as they are all
absorbed into their appropriate counter-terms. They are irrelevant
until one reaches the order at which finite size operators become
relevant, at which point renormalization is straightforward.\\

We thank Scott Hughes, Eric Poisson and Jure Zupan for helpful
comments. \uppercase{W}ork supported by \uppercase{DOE} contracts
\uppercase{DOE-ER}-40682-143 and \uppercase{DEAC02-6CH03000}.

\end{document}